\documentclass[runningheads]{cl2emult}

\usepackage{makeidx}  
\usepackage{graphicx} 
\usepackage{subeqnar} 
\usepackage{multicol} 
\usepackage{cropmark} 
\usepackage{eso}      
\makeindex            

\begin{document}
\setcounter{page}{113}
\title*{Accretion onto Black Holes and Neutron Stars:\protect\newline 
Differences and Similarities}
\toctitle{Accretion onto Black Holes and Neutron Stars: Differences
and Similarities}
%
%
\titlerunning{Accretion onto Black Holes and Neutron Stars}
%
\author{Rashid Sunyaev}
\authorrunning{Rashid Sunyaev}
%
%
\institute{Max-Planck-Institut f\"ur Astrophysik, Garching, Germany\\
\and  Space Research Institute, Moscow, Russia}

\maketitle              

\begin{abstract}
Accreting black holes and neutron stars at luminosities above 0.01 of
the critical Eddington luminosity have a lot of similarities, but also
drastic differences in their radiation and power density spectra.  The
efficiency of energy release due to accretion onto a rotating neutron
star usually is higher than in the case of a black hole.  The theory
of the spreading layer on the surface of an accreting neutron star is
discussed.  It predicts the appearance of two bright belts equidistant
from the equator.  This layer is unstable and its radiation flux
must vary with high frequencies.
\end{abstract}

\section{Introduction}

One of the most important properties of accreting black holes in our
Galaxy was discovered by Riccardo Giacconi and the {\it Uhuru} Team in
1971, when they discovered the spectral transition of Cyg X-1 from the
soft to the hard state (Tananbaum et al. 1972).  Simultaneously, a
radio source appeared in the vicinity of Cyg X-1.  Radio observations
permitted its localization with high accuracy and the identification
of the X-ray source with a bright star of the 9th magnitude.
Immediately thereafter, measurements of its optical spectrum showed
that this star is member of a 5.6-day non-eclipsing binary with an
optically invisible companion (Bolton 1972).  Lyuty et al.  (1973)
interpreted the observed ellipsoidal variations in the brightness of
the optical star as a result of the gravitational influence of a
nearby black hole invisible in optical light.  Today Cyg X-1 is the
best-known steadily accreting black hole in our Galaxy. Now we
have a list with more than 12 excellent black-hole candidates and
many of them show similar soft- to hard state transitions (Tanaka \&
Shibazaki 1996).  Recently, Cyg X-1 experienced the third transition
from a hard to a soft state in 18 years (Fig.~1).

\begin{figure}
\centering
\includegraphics[width=1\textwidth]{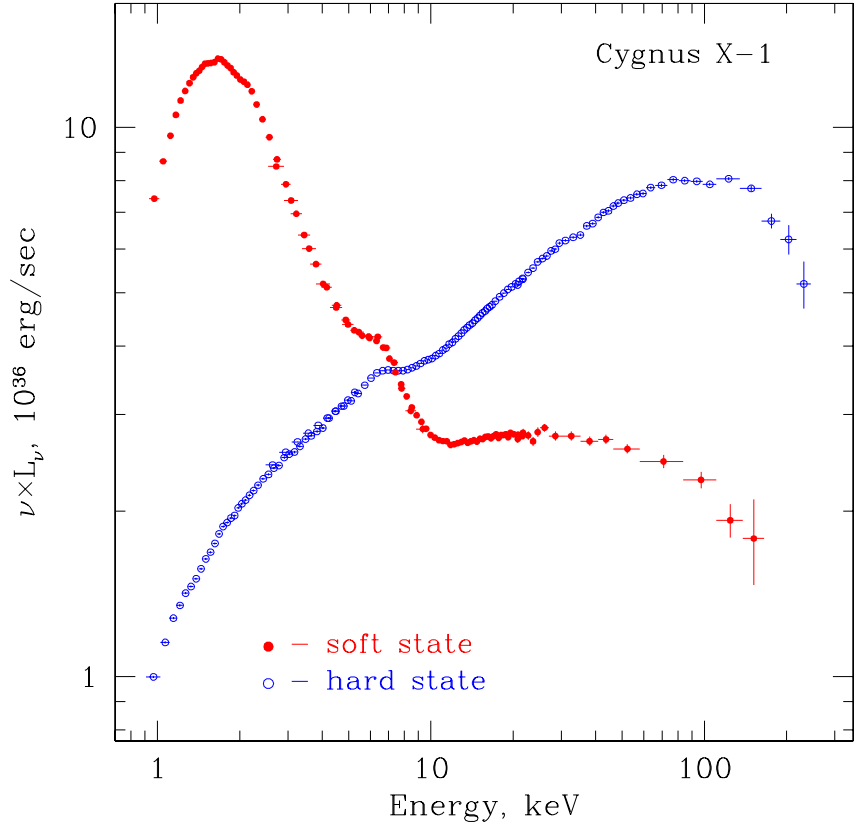}
\caption[]{The spectral energy distribution of Cyg X-1 in the soft
(filled circles) and hard (open circles) spectral state. Data are
shown of nearly simultaneous ASCA and RXTE observations on March 26,
1996 (hard state) and May 30, 1996 (soft state); cf.\ Gilfanov,
Churazov \& Revnivtsev (2000).}
\label{giacc10}
\end{figure}

Such transitions became a signature of black holes.  Today we know
that all galactic black-hole candidates show a very soft X-ray
spectrum. As predicted by standard accretion theory, this is a
multicolor disk spectrum (cf.\ Shakura \& Sunyaev 1973) or a power-law
hard X-ray spectrum with a Wien-type decay at high energies formed due
to comptonization (Sunyaev \& Tr\"umper 1979, Sunyaev \& Titarchuk
1980).  Sometimes we do not even see the high frequency decay
yet. Therefore, usually when a newly discovered X-ray transient shows
an extremely hot tail in its X-ray spectrum, we immediately refer to
it as a black-hole candidate.

Neutron stars without magnetic fields and black holes have practically
the same gravitational potential and must show many similarities.
Nevertheless, we know now that they have very different X-ray spectra
and variability characteristics.  One of the great surprises of the
last 15 years of observations is the discovery that neutron stars
also exhibit soft- to hard-state transitions (Fig.~2).

\begin{figure}
\centering
\includegraphics[width=1\textwidth]{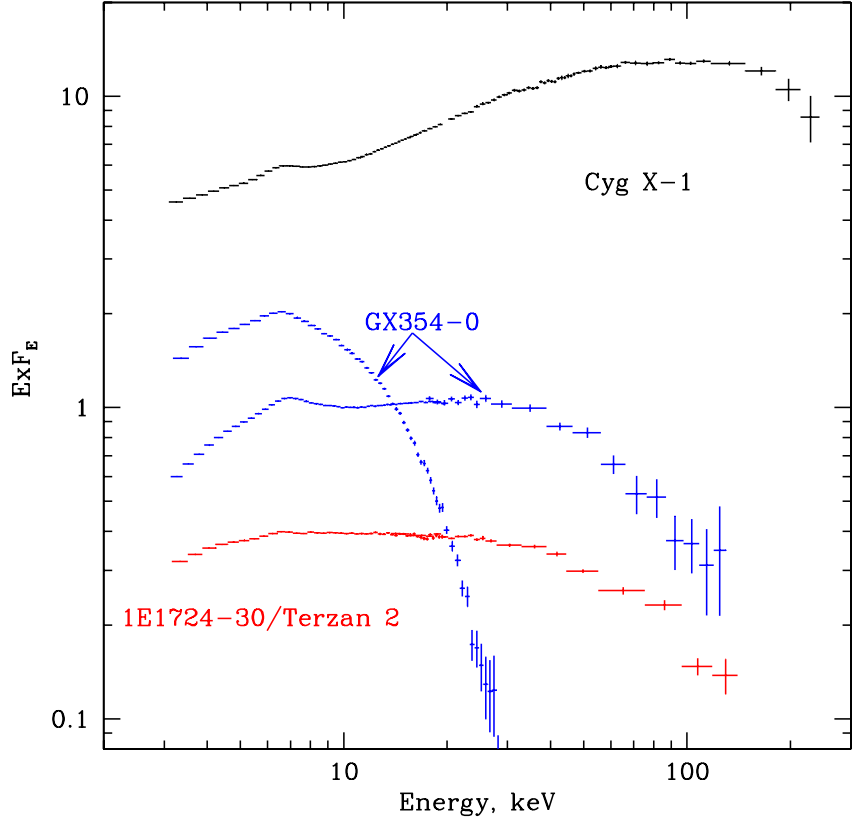}
\caption[]{The energy spectra of three X-ray binaries.  The neutron
star in GX354-0 (4U1728-34) is shown in two spectral states --
low/hard and high/soft.  Even in the hard state the neutron-star
spectra are much softer than the spectrum of Cyg~X-1 (accreting black
hole).  Adapted from Revnivtsev \& Sunyaev (2000).}
\label{giacc29}
\end{figure}
 
Neutron stars with small magnetic fields usually have spectra which
are significantly harder than the spectra of multicolor accretion
disks around black-hole candidates in a high/soft state.  But their
spectra are usually much softer than the spectra of black-hole
candidates in the hard/low state.  Sometimes we observe hot tails in
the persistent flux of X-ray bursters.  However, spectra of these hot
tails from neutron stars are much steeper than in the case of black
holes and contain a smaller fraction of the source luminosity.  It
seems that now we know the reason.  In the case of black-hole
accretion we only see the radiation of accretion disk -- plus, maybe,
the corona above it (Galeev et al. 1979) or the advection flow with
even smaller accretion efficiency (Narayan \& Yi 1995).  In the case
of neutron stars we have an object with a solid surface. Therefore,
part of the gravitational energy of the accreting matter must be
released in an extended accretion disk, and another part in the narrow
boundary layer in the vicinity of the neutron star where accreting
matter is decelerating from the Keplerian velocity (of the order of
half the velocity of light) to the velocity of rotation at the equator
of the neutron star.  The surface of the star is able to produce
enough soft protons for comptonization to cool down the hot parts of
the disk and boundary layer to temperatures below 20~keV (Sunyaev \&
Titarchuk 1989).  The physics of the boundary layer permits us to
explain the strong differences between the radiation spectra of
accreting black holes and neutron stars.  It also predicts a strong
difference in the characteristic variability timescales of the X-ray
flux from black holes and neutron stars (see below).

\section{Efficiency of Accretion onto a Rapidly Rotating Neutron Star}

The recent discovery of quasi-periodic oscillations (QPO) with
frequencies of the order of 500-600 Hz during the nuclear bursts on
the surface of a neutron star appears to be very strong evidence of
neutron-star rotation with the same frequency, or with periods of the
order of 1.6-2~ms (Strohmayer et al. 1998).  This interpretation is
natural for a nuclear burning front propagating on the surface of a
rapidly rotating neutron star.  A bright front region manifests itself
as a hot spot giving rise to the QPO.  It is important that for a
given neutron star the QPO frequency remains the same from burst to
burst.

The efficiency of accretion onto neutron stars is 
higher (usually) than the efficiency of accretion onto black holes.
The reason is obvious: in the case of a black hole we have an
event horizon and an effective energy release and the release of 
the observed radiation flux might occur only in 
the accretion flow well beyond the event horizon.  In the
case of a neutron star without a strong magnetic field 
part of the energy is released in the extended accretion
disk and another part is liberated in the narrow 
boundary layer near the surface of the neutron star.
In Newtonian mechanics energy release in the boundary
layer is equal to 
\[ L_s = {1\over 2} {GM\dot M\over R_*} (1 - {f\over f_k})^2 \, , \]  
or is equal to the energy liberated in the disk
\[ L_d = {1\over 2} {GM\dot M\over R_*}\] 
in the case of a slowly rotating compact star.   
Here and below $M$ is the gravitational mass of the star,
$R_*$  is its radius, $f_* = {1\over 2\pi} \sqrt{{GM\over R^3_*}}$
the cyclic keplerian frequency near the its surface, $f$ is
the frequency of stellar rotation and $\dot M$ is the
accretion rate.

The problem becomes much more complicated in the case of General
Relativity.  Kerr metrics is not applicable to the case of rapidly
rotating neutron star because the mass distribution within the star is
no longer spherically symmetric.  There is a strong quadrupole
component in the mass distribution.  Fortunately, there is an exact
solution of the GR equations for the case when the mass distribution
has a quadrupole component.  Using this solution, Sibgatullin \&
Sunyaev (2000) plotted the dependence of the energy release due to the
accretion onto a neutron star as a function of the rotation frequency
of that star (Fig.~3).  The existing GR solution permits us to find
the efficiency of the energy release only in the case when the spin
directions of the neutron star and accretion disk are parallel or
anti-parallel.  Unfortunately, the problem with an arbitrary angle
between the axes of rotation of the neutron star and the accretion
disk is much more complicated.  In Figure~3 the positive values of the
rotational frequency $f$ correspond to the case of corotation and
negative values describe the case of counterrotation.  The figure
presents the computations for the equation of state (EOS) FPS in the
classification of Lorenz et al. (1993) for a gravitational mass of the
neutron star $M = 1.4$~M$_\odot$.

\begin{figure}
\centering
\includegraphics[width=.7\textwidth]{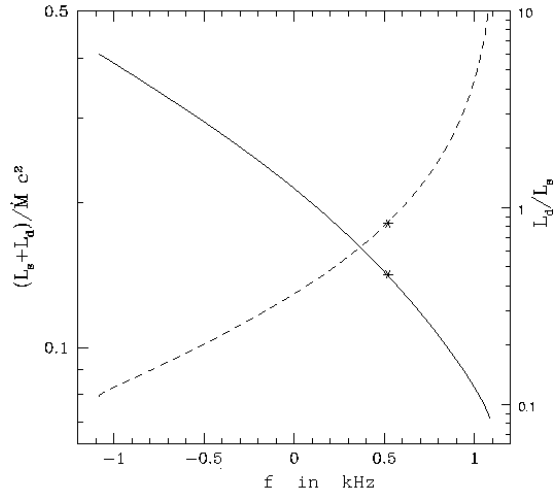}
\caption[]{Efficiency of the energy release in the disk ($L_{d}$) and
on the stellar surface ($L_{s}$) as a function of the stellar rotation
frequency $f$. Negative $f$ values correspond to the case of
counter rotation of disk and star.  The solid line and the numbers on
the left axis give the value $(L_{s}+L_{d})/\dot M c^2$.  The dashed
line and the values on the right axis give the ratio $L_{d}/L_{s}$.
A gap between the marginally stable orbit and the stellar surface
exists in the region leftside of the two asterisks on the solid and
dashed curves. There is no such gap in the case of rapid corotation of
star and disk ($f >$ 550 Hz).  Adapted from Sibgatullin \& Sunyaev
(2000).}
\label{giacc21}
\end{figure}

We see that the energy release efficiency drops rapidly with
increasing frequency in the case of corotation and increases rapidly
towards high frequencies of counter rotation.  The ratio of the disk
luminosity to the luminosity in the boundary layer or in the spreading
layer near the surface of the star also strongly depends on the
frequency of rotation.  It is close to 1 for the case of corotation
with $f = 600$~Hz and decreases up to 0.2 in the case of counter
rotation with the same frequency.

The asterisks in Figure~3 give information on the existence of a gap
between the marginally stable orbit in the accretion disk and the
radius of the star.  For frequencies of corotation higher than 550~Hz
such a gap does not exist; then the disk is in contact with the
surface of the neutron star.  For lower frequencies of corotation and
in the case of counter rotation for the EOS FPS and $M =
1.4$~M$_\odot$ there is a gap $ R_m - R_* \approx [1.44 - 3.06 (f/{\rm
kHz}) + 0.843 (f/{\rm kHz})^2 + 0.6 (f/{\rm kHz})^3 - 0.22 (f/{\rm
kHz})^4]$ km.  In the most interesting case of corotation the gap is
very narrow and the thickness of the boundary layer or the hight of
the spreading layer usually exceeds the dimension of the gap.
However, in the case of counter rotation (negative values of $f$) the
gap could be sufficiently large that it has to be taken into account.

The energy release efficiency due to accretion onto a counter-rotating
neutron star may reach very large values up to 0.67 \.{M}$c^2$ for the
case of a neutron star with baryonic mass $m = 2.1$~ M$_{\odot}$ for
$f = 1.5$~kHz and the EOS FPS.  Obviously, such a high energy release
efficiency is connected with the spin down of the rapidly (counter)
rotating star.  This efficiency is much higher than that of disk
accretion onto a Kerr black hole.  In the case of corotation the
energy release efficiency, due to accretion onto a Kerr black hole, is
higher than in the case of counterrotation.  This is reversed in the
case of accretion onto a neutron star .

\section{Structure of the Boundary Layer}

The problem of disk accretion onto a neutron star without a magnetic
field is two-dimensional.  The height of an accretion disk at low
accretion rates and luminosities ($0.01 < L/L_{\rm Edd} < 0.3$) is
small in comparison with the radius of the neutron star.  Here and
below $L_{\rm Edd} = {4\pi GMm_p \over \sigma_{T}}$ is the critical
Eddington luminosity.  The angular rotation frequency $\Omega$ in the
disk is close to keplerian and increases when matter approaches the
neutron star.  In the boundary layer the matter velocity must decrease
to the velocity of rotation at the neutron-star surface and then
matter must be redistributed over its equipotential surface.  This
surface is defined by the common influence of gravity and centrifugal
forces.  It is obvious that there must be a ring where $\Omega$
reaches its maximum, $d \Omega / dR = 0$.  There are two possible
approaches to consider the matter flow beyond this point.  We could
assume that the boundary layer is described by the same equations as
those valid for the accretion disk or we could consider the motion of
matter in the spreading layer as belonging to the surface of the
neutron star.  We tried to investigate both of these approaches in
one-dimensional approximations.  In the paper by Popham \& Sunyaev (2000)
we computed the structure and properties of the boundary layer
considering it as a part of the disk.  Figure~4 shows how the height of
the boundary layer depends on the distance from the stellar surface for
different accretion rates.  In the case of a low accretion rate or $L
\sim 0.01 \, L_{\rm Edd}$, the height of the disk in the ``neck'' between
the accretion disk and the boundary layer is close to only 40 meters and the
extension of the boundary layer about 1.5~km.  The situation
drastically changes when we go to the case of high accretion rates
with a luminosity close to the critical Eddington luminosity.  The
height of the neck between the boundary layer and the accretion disk in
this case exceeds 2~km and the boundary layer extends up to 2
neutron-star radii.

\begin{figure}
\centering
\includegraphics[width=.7\textwidth]{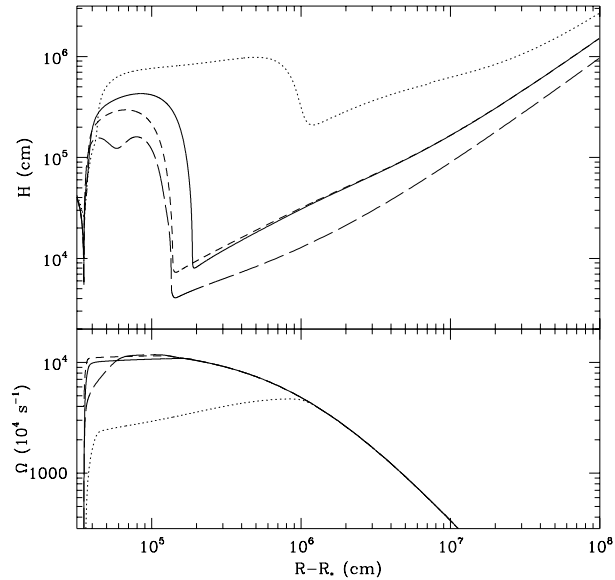}
\caption[]{The vertical pressure scale height $H$ (top) and the angular
velocity $\Omega$ (bottom), for solutions with $\dot M = 10^{-10}$
(long dash), $10^{-9}$ (solid) and $10^{-8}M_\odot yr^{-1}$
(dotted), all for a non-rotating neutron star, and for $\dot M =
10^{-9} M_\odot yr^{-1}$ and a neutron star rotation frequency $f_* 
= 636$ Hz (dashed), all with standard viscosity and $\alpha = 0.1$.
Note the very small values of $H$ at the ``neck'' between the disk 
and the boundary layer in the lower \.{M} solutions, and the rapid
increase in $H$ in the boundary layer. Adapted from Popham \&
Sunyaev (2000).}  
\label{giacc22}
\end{figure}

A more natural approach was considered by Inogamov \& Sunyaev
(1999). This approach uses the shallow water or hydraulic
approximation.  It assumes that the thickness of the spreading layer
on the surface of the neutron star is less than the circumference of
the neutron-star equator $H << 2\pi R_*$.  This approach assumes that
matter entering the equatorial ring with a very high rotational
velocity of the order of 0.5$c$, where $c$ is the velocity of
light. Then the matter begins to spiral slowly towards the poles
losing its kinetic rotation energy due to turbulent friction with the
dense underlying layer (Fig.~5 and Fig.~6).

\begin{figure}
\centering
\hbox{
\includegraphics[width=.45\textwidth]{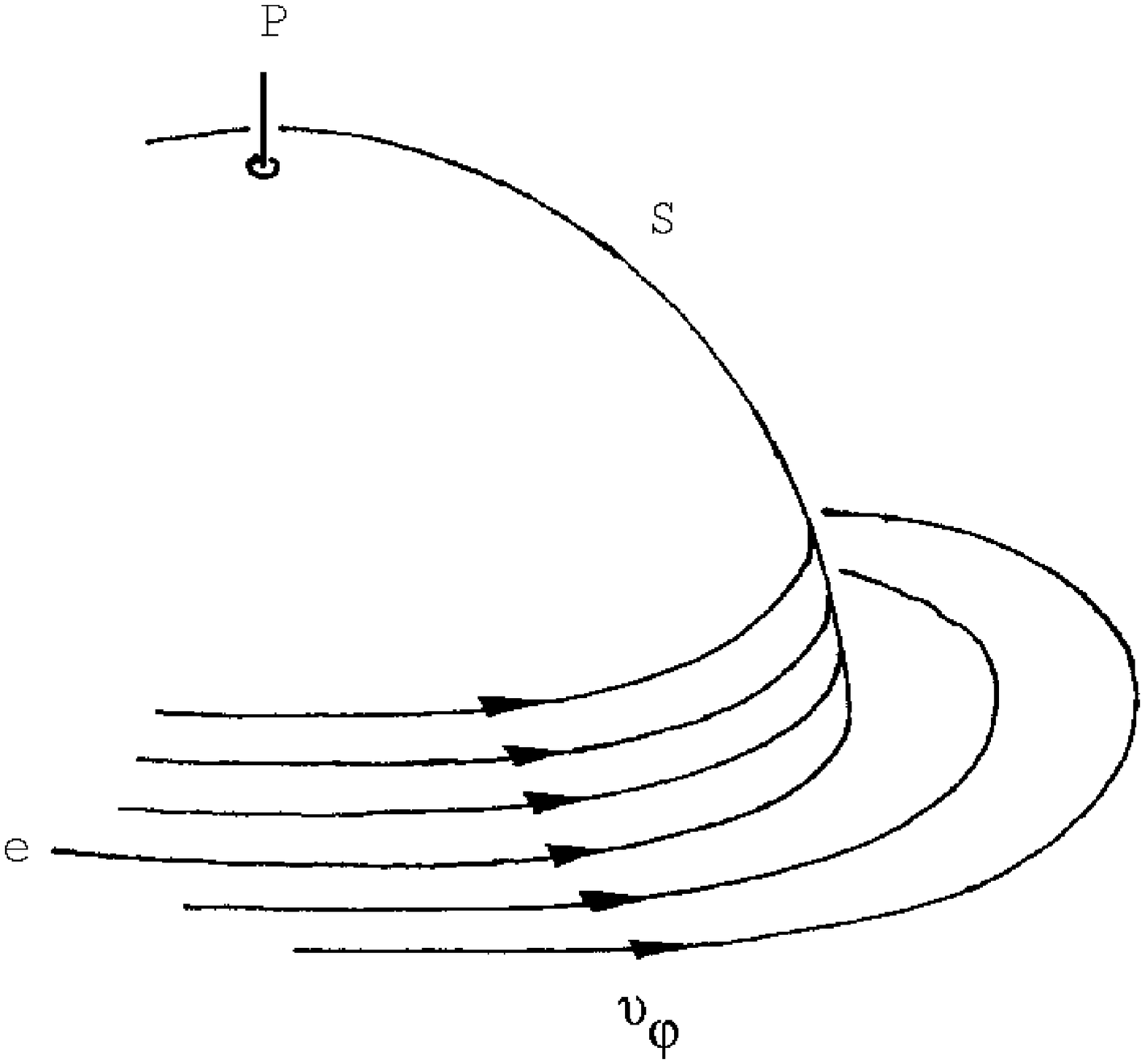}
\hspace{-2.5cm}
\vbox{
\includegraphics[width=.45\textwidth]{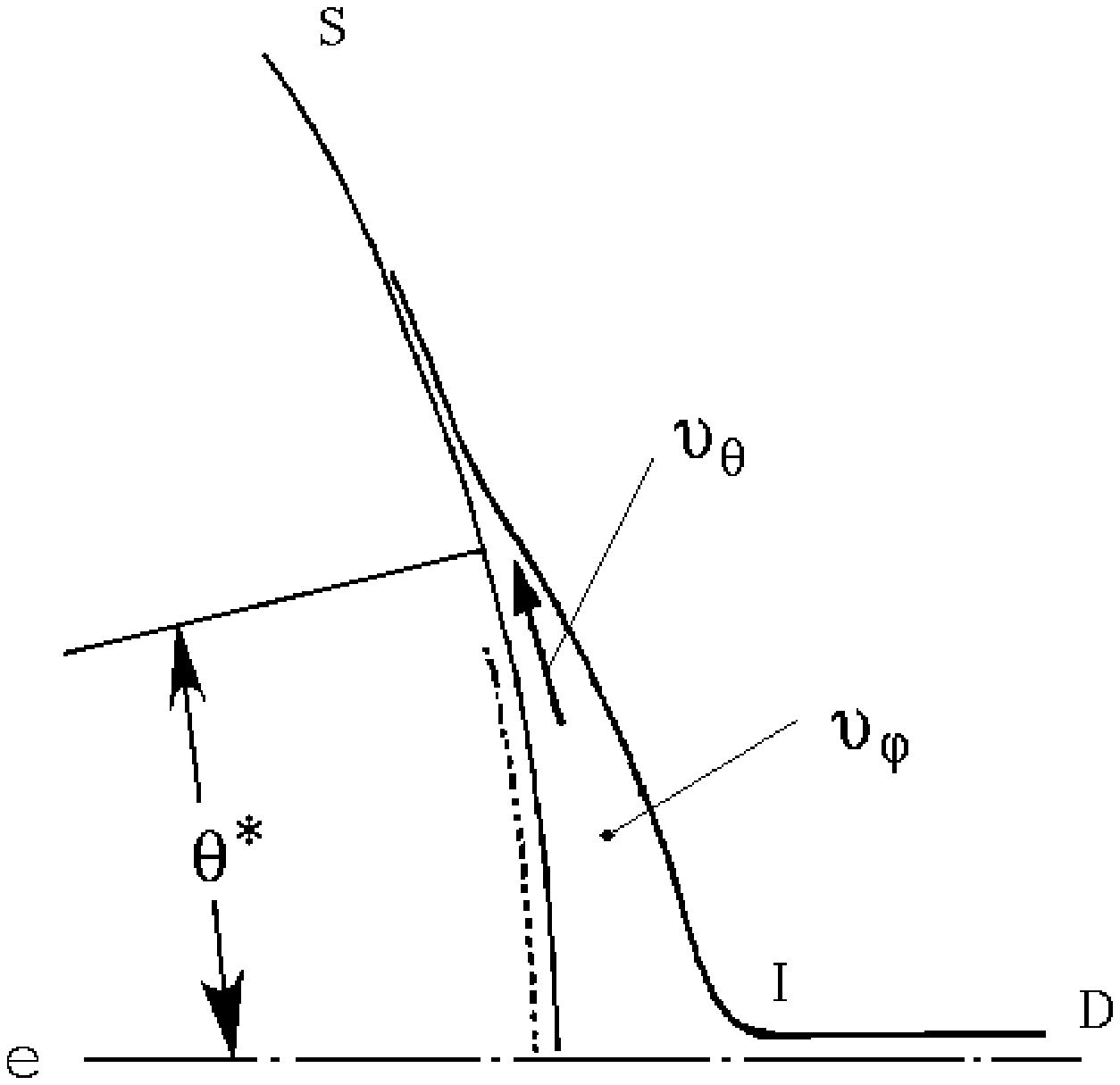}
\vspace{1cm}}
}
\caption[]{Rotation of matter in the disk and on the stellar surface
in the spreading layer model; $S$ is the stellar surface, $P$ is the
pole, and $e$ is the equator.}
\label{giacc23}
\caption[]{Spreading of the rotating plasma from the disk, $D$, over
the neutron star surface, $S$.  Here, $I$ is the intermediate zone
near the disk neck, $\theta^*$ corresponds to the position of the hot
belt, and $\theta > \theta^*$ is the cold part of the spreading
layer. The rotation velocity $v_\phi$ (filled circle) is directed
along the normal to the plane of the figure.  The slowly circulating
dense underlying layers of matter beneath the spreading layer are
indicated by the dashes. Both figures are adapted from Inogamov \&
Sunyaev (1999).}
\label{giacc24}
\end{figure}

The thickness of the spreading layer is highest in the vicinity of the
equator and decreases towards the poles.  This means that matter is
moving down the hill under the influence of gravity, the centrifugal
force and the light pressure force.  The problem is extremely
interesting.  We are dealing with radiation dominated plasma when the
radiation pressure strongly exceeds the matter pressure.  The sound
speed is close to $0.1 - 0.15 c$.  Radiative viscosity is also much
stronger than the viscosity of plasma.  The solution of the set of
hydrodynamic equations results in the following picture (see Inogamov
\& Sunyaev 1999 for details).  Two bright belts equidistant from the
equator appear on the surface of the neutron star due to disk
accretion.  Figure~7 gives the distance of the bright belts from the
equator for 4 luminosities of the neutron star: 0.01, 0.04, 0.20 and
0.80~$L_{\rm Edd}$.  The distance from the equator increases with
luminosity.

\begin{figure}
\centering
\includegraphics[width=.6\textwidth]{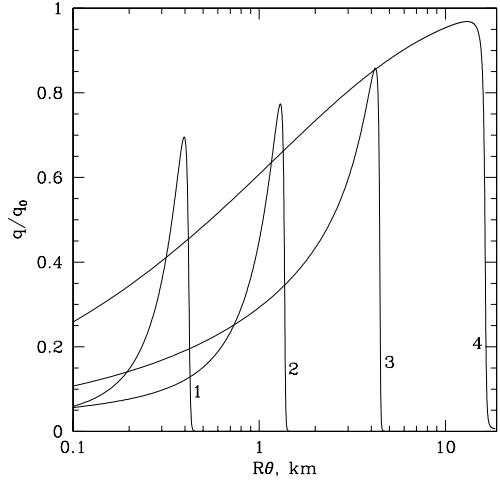}
\caption[]{Dependence of the surface brightness as a function of
distance from the equator.  Labels 1, 2, 3 and 4 refer to $L_{\rm
SL}/L_{\rm Edd} =$ 0.01, 0.04, 0.2, and 0.8, respectively; $q_o$ is
the critical Eddington flux.  Adapted from Inogamov \& and Sunyaev
(1999).}
\label{giacc25}
\end{figure}

The energy release in the vicinity of the equator is very low because
there centrifugal forces compensate gravity with high precision.
Therefore, any substantial radiation flux could destroy the structure
of the thin spreading layer.  Fortunately, advection takes the
radiation energy density and transports it to the bright belts above
and below the equator (Fig.~8).
 
\begin{figure}
\centering
\hbox{
\includegraphics[width=.5\textwidth]{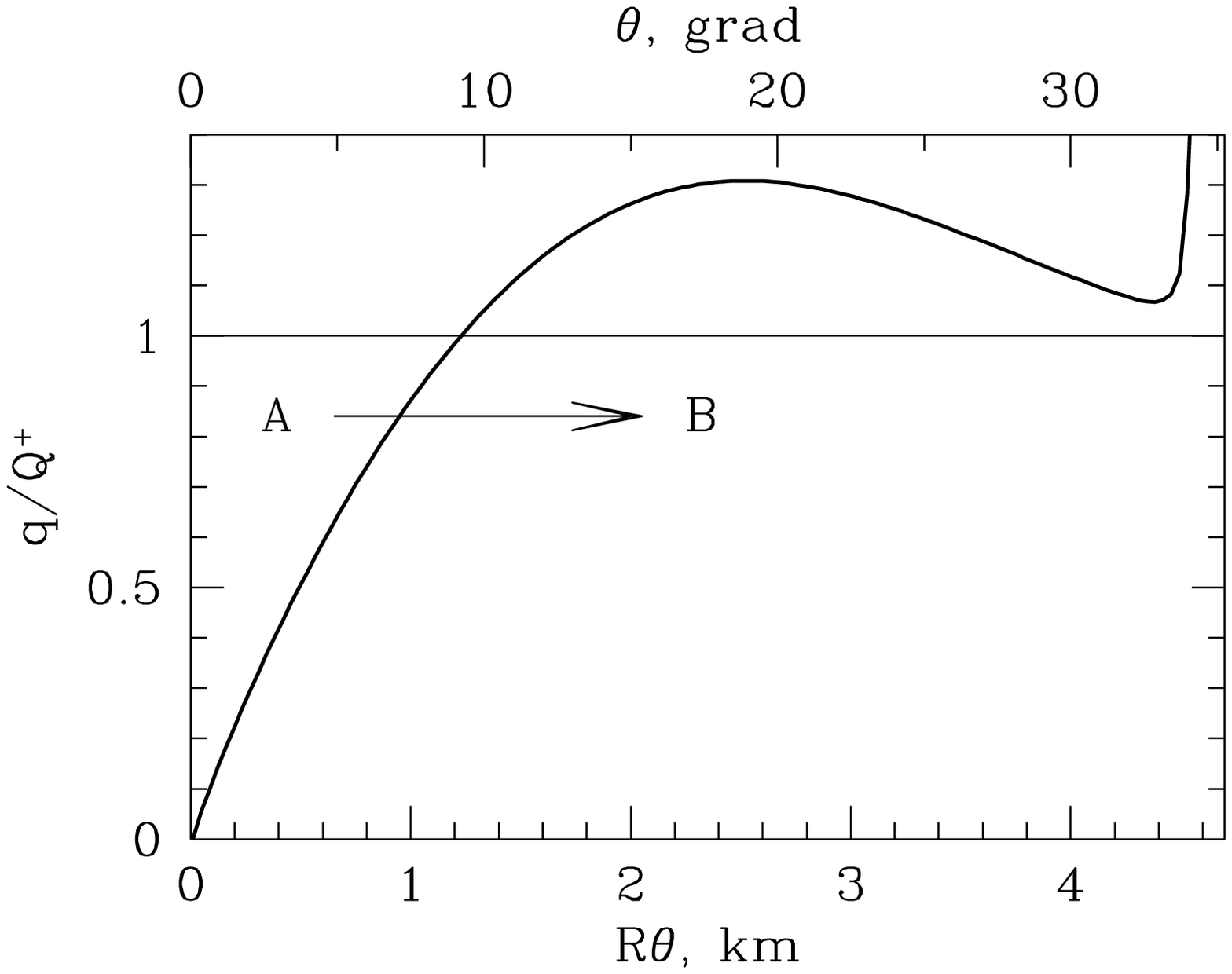}
\hspace{-3.5cm}
\vbox{
\vspace{-2.9cm}
\includegraphics[width=.5\textwidth]{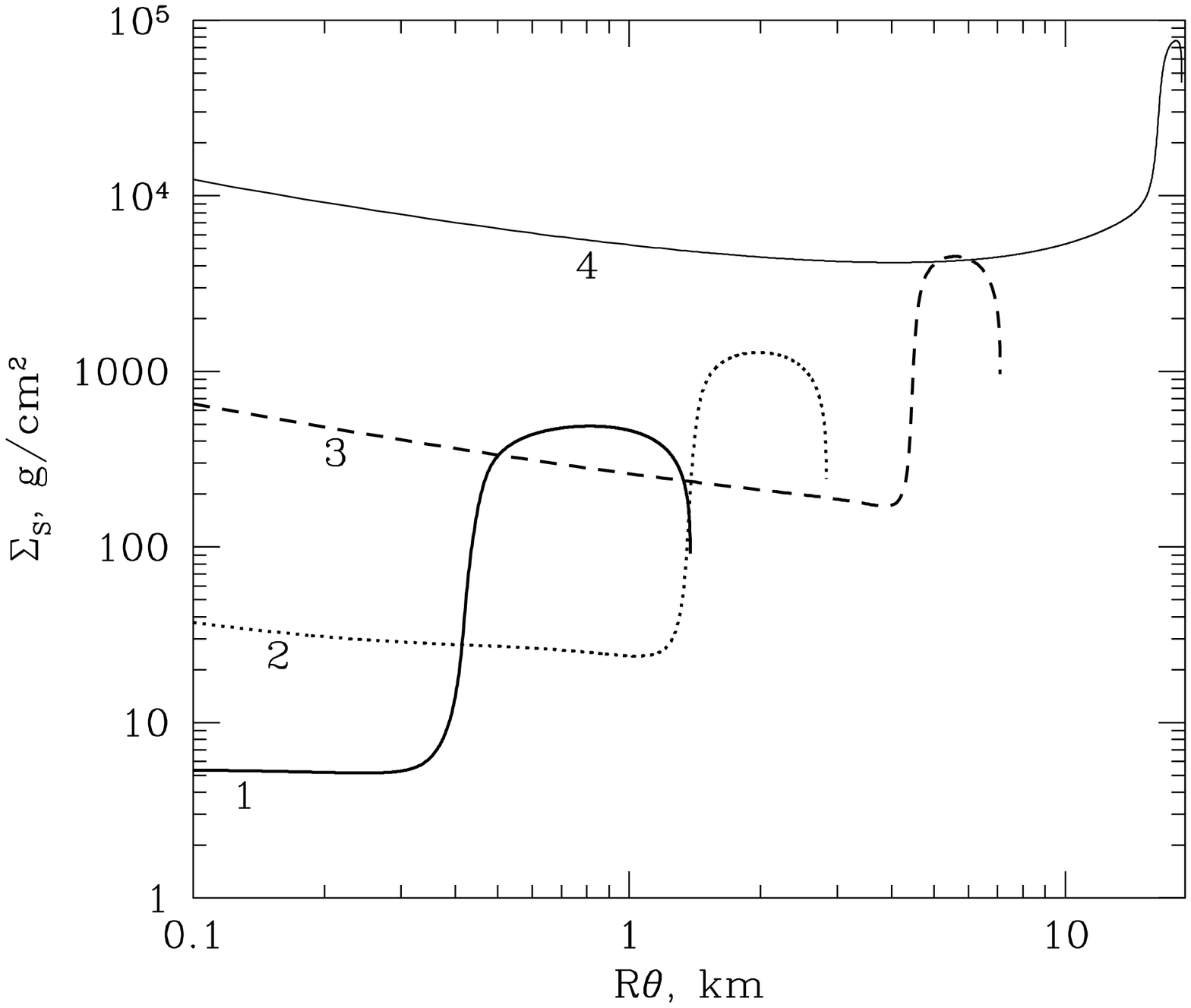}}
}
\caption[]{The dynamics of the spreading layer is determined in many
ways by the hydrodynamic transfer of radiative energy.  The ratio of
the energy flux, $q$, emitted per unit area of the radiating layer to
the frictional energy release $Q^+$ per unit area of the contact
surface between the spreading layer and the star.  The energy is
transferred from zone A into zone B through meridional advection.
This calculation is for $L_{\rm SL}/L_{\rm Edd} = 0.2$.}
\label{giacc26}
\caption[]{Column density of matter in the spreading layer
as a function of the distance from the equator.  A strong
increase of $\Sigma$ occurs when the flow cools down and
begins to move very slowly.  Bright belts correspond to the
regions of minimal $\Sigma$.  Labels 1, 2, 3 and 4 refer to
$L_{\rm SL}/L_{\rm Edd} =$ 0.01, 0.04, 0.2 and 0.8, respectively.
Both figures are adapted from Inogamov \& Sunyaev (1999).}
\label{giacc27}
\end{figure}

In these bright belts the rotational velocity of the spreading matter
becomes low enough to permit the existence of a large radiation flux
comparable to the critical Eddington flux $q_{0} = {m_{p} c^{3} \over
2 \sigma_{T}R_g} ({R_g\over R_*})^2 = 10^{22} {{\rm W}\over {\rm m}^2}$,
where $R_g$ is the gravitational radius.  This flux value is
comparable to radiation fluxes achieved in the most intense petawatt
laser facilities (Perry 1996, Budil et al. 2000).

We are dealing here with a critical Eddington flux even in the case of
a low luminosity of the neutron star ($0.01 < L/L_{\rm Edd} < 1$).  The
surface of the bright belts is small and the high radiation flux from
the narrow belts is consistent with the low luminosity of the star.

The matter in the spreading layer is practically levitating.  The
difference between the gravitational force and the centrifugal- and
radiation pressure force is close to $(1-3)\times 10^{-3}$ of gravity.
At higher longitudes the rotational velocity of matter and the
velocity of the flow along the meridian decreases and the flow becomes
subsonic, cool, dense and very slow.

One of the most interesting predictions of the theory of the spreading
layer is the strong dependence of the matter column density in the
spreading layer on the accretion rate or the luminosity of the neutron
star (see Fig.~9).  In the case of a low luminosity the levitating
layer in the bright belts is optically thin against Thompson
scattering $\tau_{T} \sim 2$.  Under these circumstances it is
impossible to radiate the energy released due to accretion at low
temperatures.  Comptonization forms hard tails.  In the case of a high
luminosity the bright belt has a large column density (up to
10~kg/cm$^2$).  Then free-free processes and comptonization form
Bose-Einstein type spectra inside the spreading layer and the
resulting spectrum is much softer than in the case of low luminosity.

\section{Time Variability in the Accretion Disk\protect\newline 
and in the Boundary Layer}

All instabilities existing in the accretion disk modulate the flow of
matter onto the neutron-star surface.  Therefore, we could expect that
the majority of the types of variability we observe in accreting black
holes must manifest themselves in accreting neutron stars with
characteristic timescales proportional to the mass of the accreting
object (see e.g. Shakura \& Sunyaev 1976, Wijnands \& van der Klis
1999).  The spreading layer on the surface of the neutron star is the
source of additional high-frequency instabilities (see the discussion
in Sunyaev \& Revnivtsev 2000).  Their origin is obvious -- the matter
in the bright belts is radiation dominated, levitating, the height is
smaller than in the region of the main energy release in the accretion
disk, the sound velocity is huge and corresponding sound frequencies
are very high.  

\begin{figure}
\centering
\includegraphics[width=.9\textwidth]{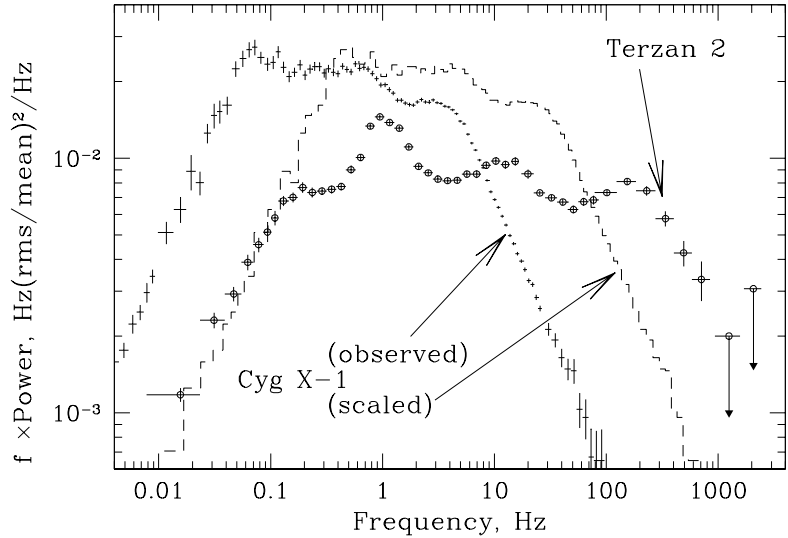}
\caption[]{Comparison of the power spectra of a black hole (Cyg~X-1)
and a neutron star (Terzan~2).  The dashed line shows the power
spectrum of Cyg X-1 scaled according to the mass ratio with Terzan 2
($f_{\rm Cyg X-1} \times 7 \rightarrow f^{\rm scaled}_{\rm Cyg X-1}$.
This simple scaling is important but insufficient to explain fully the
difference in the high frequency variability of Terzan`2 and Cyg~X-1.
Note that the slopes of the power density spectra of Terzan~2 and
Cyg~X-1 in the high and low frequency limits are similar, but that the
power spectrum of Terzan 2 is significantly broader than that of
Cyg~X-1.  Adapted from Sunyaev \& Revnivtsev (2000).}
\label{giacc28}
\end{figure}

Sunyaev \& Revnivtsev (2000) compared the power
density spectra of 9 black holes and 9 neutron stars observed by RXTE
in their low/hard state.  There is a very strong difference.  In the
power density spectra of accreting neutron stars with a weak magnetic
field significant power is contained at frequencies close to one kHz.
At the same time, most Galactic accreting black holes demonstrate a
strong decline in the power spectra at the frequencies higher than
10-50~Hz.  In principle this might open an additional way to
distinguish the accreting neutron stars from black holes in X-ray
transients (we do not mention in this paper the well-known
differences: X-ray bursts or X-ray pulsations). Fig.~10 compares the
power density spectrum of Cyg~X-1 with the power density spectrum of
the X-ray burster in Terzan~2.

The simplest assumption is that the characteristic frequencies in the
power spectra of the sources scale as M$^{-1}$ (Shakura \& Sunyaev
1976).  This scaling law is valid for e.g. the keplerian frequency in
the vicinity of the marginally stable orbit, the thermal and secular
instabilities of the accretion disk in the region of main energy
release, and the Balbus-Hawley instability.  However, this assumption
does not account for the observed difference in the high frequency
variability between neutron stars and black holes.

\end{document}